\documentclass[prl,twocolumn,nofootinbib,floatfix,sort&compress,letterpaper,numbers,superscriptaddress]{revtex4}
\pdfoutput=1
\usepackage{amsmath,amsfonts,amssymb}
\usepackage{graphicx}
\usepackage{hyperref}
\usepackage{verbatim}
\usepackage{mathrsfs}
\usepackage{natbib}
\usepackage{braket}
\usepackage{color}

\newcommand{\bi}{\begin{itemize}} \newcommand{\ei}{\end{itemize}}
 
\newcommand{\bn}{\begin{enumerate}} \newcommand{\en}{\end{enumerate}}
\newcommand{\ba}{\begin{array}} \newcommand{\ea}{\end{array}}
\newcommand{\bc}{}
\newcommand{\be}{\begin{equation}} \newcommand{\ee}{\end{equation}}
\newcommand{\bex}{\begin{equation*}} \newcommand{\eex}{\end{equation*}}
\newcommand{\bea}{\begin{eqnarray}} \newcommand{\eea}{\end{eqnarray}}
\newcommand{\beax}{\begin{eqnarray*}} \newcommand{\eeax}{\end{eqnarray*}}

\newcommand{\btp}[6]{\begin{tikzpicture}[smooth, domain=#2:#3]  \draw[->](#2  ,0) -- (#3,0 ) node[right]{#1}; \draw[->](0,#5) -- (0,#6) node[above] {#4}; \clip (#2,#5) rectangle(#3,#6);}

\newcommand{\dd}{\mathrm{d}}

\begin{document}
\renewcommand*{\thefootnote}{\fnsymbol{footnote}}

\title{Aspects of nucleation on curved and flat surfaces}
\author{Eric M. Horsley}
\affiliation{Department of Physics and Astronomy, University of Pennsylvania, Philadelphia, Pennsylvania 19104, USA}
\author{Maxim O. Lavrentovich}
\affiliation{Department of Physics and Astronomy, University of Tennessee, Knoxville, Tennessee 37996, USA}
\author{Randall D. Kamien\thanks{Electronic address: kamien@upenn.edu}}
\affiliation{Department of Physics and Astronomy, University of Pennsylvania, Philadelphia, Pennsylvania 19104, USA}

\date{\today
}

\begin{abstract}
We investigate the energetics of droplets sourced by the thermal fluctuations in a system undergoing a first-order transition. In particular, we confine our studies to two dimensions with explicit calulations in the plane and on the sphere.  Using an isoperimetric inequality from the differential geometry literature and a theorem on the inequality's saturation, we show how geometry informs the critical droplet size and shape. This inequality establishes a ``mean field" result for nucleated droplets.  We then study the effects of fluctuations on the interfaces of droplets in two dimensions, treating the droplet interface as a fluctuating line. We emphasize that   care is needed in deriving the line curvature energy from the Landau-Ginzburg energy functional and in interpreting the scalings of the nucleation rate with the size of the droplet.   We end with a comparison of nucleation in the plane and on a sphere.
\end{abstract}

\maketitle

We are familiar with nucleation in  first-order phase transitions where a system in a particular phase becomes unstable to the spatially-localized nucleation and growth of a new, more favorable phase.  Some common examples include the condensation of a vapor into liquid droplets or the freezing of a liquid upon cooling. We will be interested in the scenario where the preexisting phase is metastable, and the new, stable phase can only form if thermal fluctuations can overcome the energetic barrier associated with the formation of a nucleus of the stable phase, which may then grow.   Although these familiar processes typically occur in three dimensions, there are many naturally-occurring and engineered systems which have ordered phases nucleate and grow on surfaces. Examples of such phase transitions include crystallization of glucose isomerase in the plane \cite{Sleutel2015}, colloidal crystal assembly at a curved oil-water interface \cite{manoharan}, phase separation in a lipid bilayer vesicle \cite{baumgart}, and the ordering of a block copolymer film deposited on a curved substrate \cite{copolymerfilm}. These processes may also describe aspects of biological processes such as viral capsid \cite{Zandi2006,capsid1,capsid2} and pollen grain assembly \cite{pollen}.  Luque {\it et al.} have studied the growth of shells from identical subunits on the sphere and how their behavior is controlled by an effective line tension\cite{Luque2012}.   Finally, the processes discussed here are also relevant for the study of the fate of false vacua (metastable states in the language we adopt here) in cosmology \citep{Coleman1977,Coleman1977erratum,Callan1977}. 

  In our analysis we will always consider systems in some ``disordered'' state driven ({\sl e.g.}, by cooling) to a region of parameter space where the ``ordered state'' becomes more energetically favorable, but the disordered state remains metastable. In this context, fluctuations will drive the appearance of nuclei of the ordered state.  One of the most important observables to calculate is the nucleation rate $\Gamma$, which we may argue on quite general grounds \cite{Langer1967} has the form
\begin{align}
\Gamma=\Gamma_0e^{-\beta E^*}, \label{eq:nucrate}
\end{align}
originally derived by Kramers in the context of diffusive escape over a potential barrier \cite{Kramers}.  Here, $\beta\equiv (k_B T)^{-1}$ is the inverse temperature, $E^*$ is the difference between the energy of the metastable state and the critical nucleated cluster of the ordered state, and  $\Gamma_0$ is a prefactor which, in a chemical system, is derived from the microscopic kinetics. Within a field theoretic view, it is possible to compute the rate $\Gamma$ directly from the imaginary part of the appropriate free energy.  This is particularly surprising since nucleation is an essentially non-equilibrium process and the free energy used comes from doing an equilibrium calculation.  The theoretical foundation for the form of these nucleation rates was developed by Langer in a series of papers \citep{Langer1967,Langer1968,Langer1969}.  Furthermore, factors in $\Gamma_0$ scaling with the size of the droplet were found and identified as universal insofar as they did not depend on underlying model parameters \cite{Gunther1980,Langer1967}.

The focus of our study will be the nucleated cluster energy  $E$. In many cases such as the pollen and copolymers, the ordered phases have interesting spatial structure which may have important consequences for the nucleation processes, such as the presence of anisotropic surface tensions in the nuclei \cite{anisotropictension}. In addition, it must be noted that for crystals nucleating on curved surfaces there are additional elastic effects arising from geometric frustration which at certain length scales in the nucleation process are non-negligible \cite{Meng2014,Grason2016}. For simplicity, we ignore the fine structure of the ordered phase and associated energetic contributions. Then, within classical nucleation theory (CNT), the energy of a single nucleated cluster takes the phenomenological form
\begin{align}
E=\gamma P - c A, \label{eq:dH}
\end{align}
where $\gamma$ is the line tension, $P$ is the perimeter of the cluster, $c$ is the difference in the bulk energy density of the metastable state and ground state, and $A$ is the area of the cluster. The physical variables here are chosen to correspond to the two-dimensional problem, but the original development of CNT would refer to the three-dimensional analogs (see \cite{Kalikmanov2013} for a review of CNT and the relevant literature). There are two issues to address in arriving at an improved understanding and expression of \eqref{eq:dH}: the possible   dependence of the line tension $\gamma$ on the cluster size and the consequences of a fluctuating interface.

Tolman was one of the first to address the possibility of the size dependence of the surface tension  \cite{Tolman1949}, seeming to resolve the issue (see the early references in Tolman's paper for the relevant discussions). Tolman's result for the surface tension $\sigma$ of a nucleated cluster in three dimensions reads
\begin{align}
\sigma \equiv \sigma(R)=\sigma_{\infty}\left(1-\frac{2\delta_T}{R}\right),
\end{align}
where $R$ is the radius of the cluster, $\sigma_{\infty}$ is the surface tensions of the infinite, flat interface, and $\delta_T$ is the Tolman length.  The Tolman length is generally small, nearing the scale of the molecules themselves, and therefore the correction only becomes significant for very small clusters. 
    However, experimental work looking into the size-dependence of the surface tension has found conflicting results (see the work of Bruot and Caupin \cite{Bruot2016} and references therein).  These more recent works have not agreed on the sign of the Tolman length, and it has been suggested by Bruot and Caupin that future work should consider higher order corrections in $1/R$.  Despite over half a century of work, robust conclusions seem few and far between, and it may be valuable to reexamine Tolman's arguments and conclusions \citep{Horsch2017}.  We will explore the fate of the Tolman length for nucleation on  two-dimensional surfaces. 
 
 The impact of fluctuations of an interface within a field theoretic context was touched upon by Langer \citep{Langer1967}, but received a full treatment by G\"{u}nther {\sl et al} \citep{Gunther1980}. In this context, one shows that an effective membrane energy is achieved from a Landau-Ginzburg functional of the order parameter. Specifically, the surface tension and bending moduli appearing in the effective membrane energy are related to  the derivatives of the soliton solutions of the Euler-Lagrange equations for the order parameter. It is also possible to perform a similar analysis in a curved background \citep{Garriga1994}.   Our work considers a similar  Landau-Ginzburg functional for a scalar order parameter in curved and flat geometries, which we use to derive an effective membrane energy reminiscent of the Canham-Helfrich Hamiltonian on both the plane and the sphere. We then use a more geometrical approach similar to Voloshin's analysis of nucleation rates on the plane \cite{Voloshin1985} (rather than the functional analysis approaches of \cite{Gunther1980,Garriga1994}) to study the nucleated droplet shape and the thermal fluctuations in curved and flat backgrounds.

The paper is organized as follows. In the first section, we consider the critical radius $R^*$ for a nucleating droplet, where the critical condition is that the energy $E(R)$ of the droplet satisfies  $\left.dE/dR \right|_{R=R^*}=0$.  We can ask if the critical droplet radius $R^*$ is smaller or larger as we change the background curvature.  This question was addressed by G\'omez et al. \cite{Gomez2015}.  We show that these results follow from a reasonable phenomenological model and a general, yet simple, isoperimetric inequality. Furthermore, this result acts as a sort of mean field around which we include interface fluctuations.

In the next section we show schematically how one can relate the original CNT form of the energy \eqref{eq:dH} with that of the Landau-Ginzburg functional evaluated at soliton solutions of the Euler-Lagrange equations. In addition, we provide the results of the Tolman calculation for the size dependence of the line tension in two dimensions. These results help in the  interpretation of later results which include fluctuations.

In the final section,  we derive an effective interface energy to look for corrections to \eqref{eq:dH} due to thermal fluctuations. Our general approach is analogous to the three-dimensional analysis of Prestipino, Laio, and Tosatti \citep{Prestipino2012,Prestipino2013,Prestipino2014} and the two-dimensional analysis of Voloshin \cite{Voloshin1985}, except that we  extend the analysis to nucleation on curved surfaces and discuss different regularizations of the high energy fluctuating modes. We conclude with a discussion of possible future work and the implications of our results for CNT.

\section{An inequality and a theorem}

The nucleation droplet shape with a fixed area will naturally minimize the contribution from the perimeter. We'll assume for now that this perimeter has an infinitesimal width and we will ignore any spatial variation of the order parameter. We can than think about the nucleation droplet phenomenologically, as an area $A$ with some perimeter of length $P$ living on some surface.  Then, we may make use of isoperimetric inequalties relating the size of an object to the size of its boundary.  The following inequality derived by Morgan, Hutchings, and Howards \citep{Morgan2000} fits the bill: 
\begin{equation}
P^2 \geq 4 \pi (\chi-f+1)A - 2 \int_0^A G(t)\,\mathrm{d}t, \label{eq:isoineq}
\end{equation}
where  $f$ is the number of components of the droplets, $\chi$ is the Euler characteristic, and $G(t)$ is the supremum of the total Gaussian curvature for a given region area $t$.  It would be nice if we could saturate the inequality in Eq.~\eqref{eq:isoineq} to find regions which minimize the perimeter $P$ for a given area $A$. Thankfully, for a reasonably large set of surfaces, such a saturation is possible via the following theorem:

\

\noindent \textbf{Isoperimetric Theorem\citep{Morgan2000}.} \textit{Consider a plane, sphere, real projective plane, or closed disk $S$ with smooth, rotationally symmetric metric such that the Gauss curvature is a nonincreasing function of the distance from the origin. Then among disjoint unions of embedded disks of a given area, a round disk centered at the origin minimizes perimeter.\\
\indent It is unique, except of course that a circle in the interior of a ball about the origin
of constant Gauss curvature may be replaced by a congruent circle in that ball.}

\

    Therefore, for a surface with a constant Gaussian curvature, we know that perimeter-minimizing area is a disk and that its perimeter $P$ satisfies, via  Eq.~\eqref{eq:isoineq}, 
\begin{equation}
P^2= 4 \pi A-K A^2, \label{eq:perim}
\end{equation}
where $K$ is the constant Gaussian curvature of the surface. For a sphere, we would have $K=1/R_s^2$, with $R_s$ the sphere radius. We'll now suppose that we have a line tension $\gamma$ penalizing the perimeter and a condensation energy  area density $c$. Then,  the phenomenological form in Eq.~\eqref{eq:dH} combined with Eq.~\eqref{eq:perim} yields
\begin{equation}E = \gamma \sqrt{4 \pi A-K A^2} - cA.\end{equation}
To find the critical droplet area, we have to specify how the droplet would evolve in time.  Perhaps the simplest choice of the dynamics is that the droplet area will grow or shrink, driving the energy $E$ to a minimum. These relaxation dynamics are: $\partial_t A=- \omega (dE/dA)$, where $\omega$ is inversely proportional to a characteristic relaxation time.  We would then set $\partial_t A=0$ to find the critical area $A^*$ at which the seed starts to grow.  We find
\begin{equation}
A^*= \frac{2 \pi}{K} \left[ 1 - \frac{c}{\sqrt{c^2+K \gamma^2}} \right].
\end{equation} 
This formula also works for nucleation on a flat plane: Taking the $K \rightarrow 0$ limit, we find $A^* \rightarrow \pi \gamma^2/c^2$. Also, the critical area $A^*$ may be related to the critical radius $R^*$ of the geodesic disc, since $A = 2 \pi[1-\cos (\sqrt{K}R)]/K$, where $R$ is the geodesic distance from the center of the droplet to the edge.  We find that the critical droplet radius is given by
\begin{equation}
R^*= \frac{1}{\sqrt{K}} \arctan \left( \frac{\sqrt{K} \gamma}{c} \right),
\end{equation}
recapitulating the result in Eq.~11 in Ref.~\cite{Gomez2015}. Let us take these mathematical results as our foundation and build up the theory of droplets in two dimensions.
\section{Thermodynamic potentials and Landau-Ginzburg energies}
We will now consider possible corrections to the phenomenological energy in Eq.~\eqref{eq:dH}. First, we consider the possibility that the line tension $\gamma$ depends on the shape of the droplet.  To begin, we will follow Tolman's classical thermodynamic analysis \cite{Tolman1949}, but work in two dimensions.  This analysis consists of writing down the change in free energy of the combined liquid-vapor system giving the usual expression for the Laplace pressure.  Then, using the Gibbs adsorption equation and the Laplace pressure, one finds a relation between the line tension $\gamma$ and the radius $R$ of the nucleated drop:
\begin{align}
\mathrm{d} \gamma= - \frac{\Upsilon}{\Delta \rho} \, \mathrm{d} \left( \frac{\gamma}{R}\right), \label{eq:Tolmanrelation}
\end{align}
where $\Upsilon$ is the order parameter ({\sl e.g.}, mass density) computed at the interface ({\sl i.e.}, a per-unit-length density) and $\Delta \rho$ is the change in order parameter between the ordered and disordered state ({\sl i.e.}, a per-unit-area density).  Recapitulating Tolman's arguments, one finds that the ratio $\Upsilon/\Delta \rho$  may be parameterized in terms of a length $\delta$ as follows: $\Upsilon/\Delta \rho = \delta+\delta^2/(2R) $. Substituting this parameterization into Eq.~\eqref{eq:Tolmanrelation} and integrating yields 
\begin{align}
 \gamma(R)&=\gamma_{\infty}\frac{ e^{\frac{\pi}{4}-\tan^{-1}\left(1+\frac{\delta}{R}\right)}}{\sqrt{1+\frac{\delta}{R}+\frac{\delta^2}{2R^2}}}\\
&\approx \gamma_{\infty}\left(1-\frac{\delta}{R}+\mathcal{O}\left(R^{-2}\right)\right),
\end{align}
where we may now identify the $\delta$ as a length associated with the first $1/R$ correction to the line tension. This is the so-called Tolman length, which has evidently the same character in two and three dimensions, at least within  Tolman's original thermodynamic analysis.  We have therefore identified a potential correction to our theory.

\begin{figure}
\includegraphics[width=\linewidth]{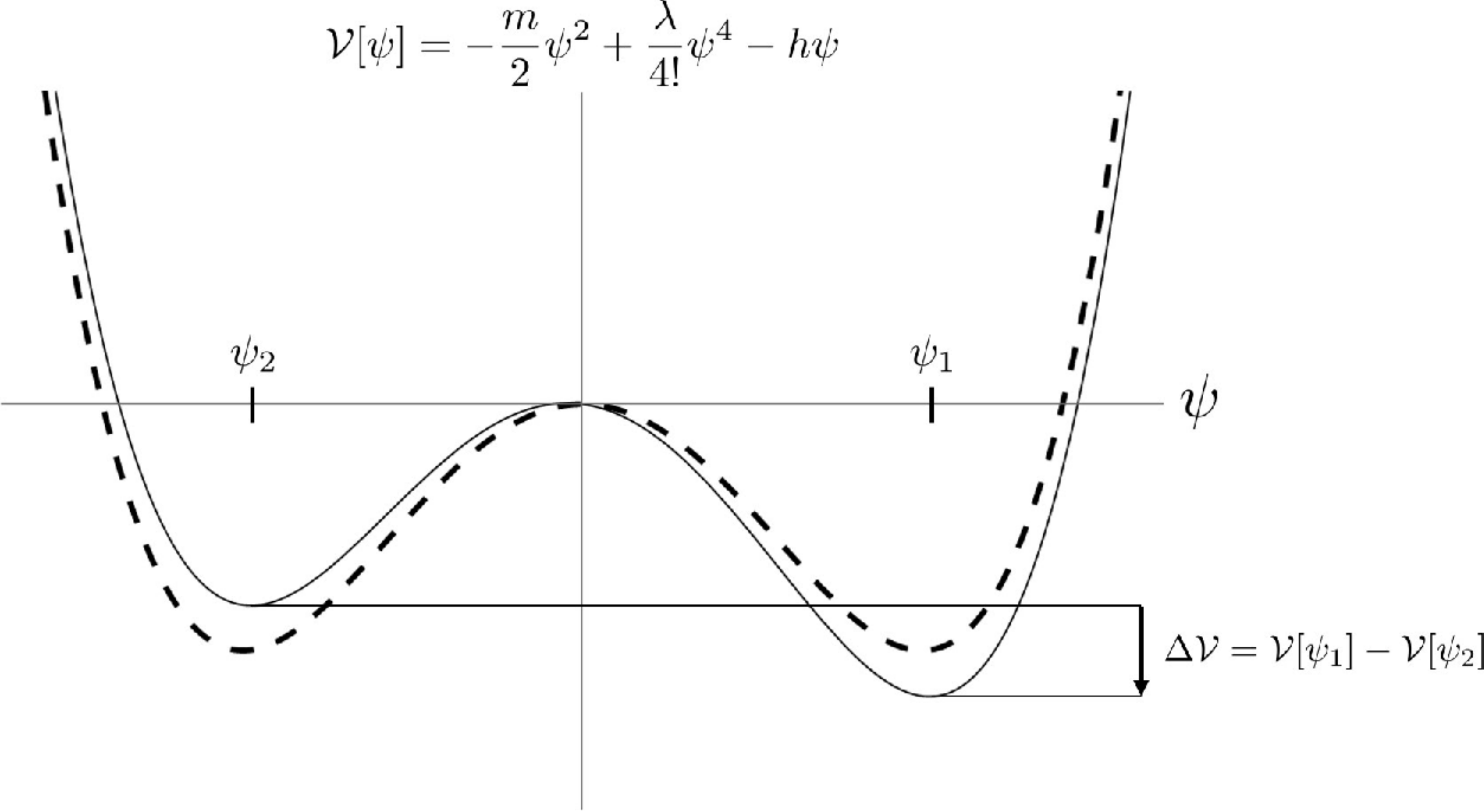}
\caption{The dashed curve corresponds to the scenario where the minima are degenerate $h=0$. The solid curve depicts the development of a metastable and stable state, $\psi_2$ and $\psi_1$, respectively.  For both curves $m>0$ and $\lambda > 0$.}
\label{fig:asympotential}
\end{figure}

We have until now considered droplets with infinitely sharp and static interfaces. However, realistic nuclei will have finite thickness and  fluctuating interfaces.  We may therefore consider a  scalar order parameter $\psi(x)$ that can capture spatial variation in the interface between phases. The field $\psi(x)$ may represent the local density of material or perhaps the degree of crystallinity.   We will suppose that $\psi(x)$ takes on constant ({\sl i.e.}, spatially uniform) values $\psi_1>0$ in the ordered phase and $\psi_2<0$ in the disordered phase (see Fig.~\ref{fig:asympotential}).  Since we will be interested in first-order transitions between these phases, we can set up a potential for $\psi(x)$ that has two potential wells at $\psi_{1,2}$, separated by an energy barrier.  A simple form for that potential is
 \begin{align}
 \mathcal{V}(\psi)=-\frac{m\psi^2}{2}+\frac{\lambda \psi^4}{4!}-h\psi, \label{eq:potential}
 \end{align}
   where $m,\lambda>0$ are phenomenological parameters and $h$ is a bias that we can tune to make the ordered or disordered phase more energetically favorable by setting $h>0$ or $h<0$, respectively. At coexistence when the two phases are equally favorable and $h=0$, we readily find that $\psi_{1,2}= \pm \sqrt{6 m/\lambda}$ for the potential parameterization in Eq.~\eqref{eq:potential}.  Penalizing spatial variations in $\psi(x)$,  the free energy functional $\mathcal{H}$ associated with the field $\psi(x)$ may be written as
\begin{align}
\mathcal{H}&=\int \dd^dx\left(\frac{\kappa}{2}(\nabla \psi)^2+\mathcal{V}(\psi) \right).\label{energyfunc}
\end{align}
 In the absence of thermal fluctuations, the equilibrium configuration of the field $\psi$ will extremize the functional $\mathcal{H}$.  Such a saddle-point solution will necessarily satisfy the Euler-Lagrange equation\begin{align}
\frac{\delta \mathcal{H}}{\delta \psi}=0=-\kappa\nabla^2{\psi} +\frac{\partial\mathcal{V}}{\partial \psi}. \label{ELequ}
\end{align}
Two obvious solutions to this equation are the spatially uniform states $\psi = \psi_{1,2}$.  However, when $h=0$, we also have a different solution $\psi \equiv \psi_I(z)$  that varies along one direction $z$: 
\begin{align}
{\psi}_I(z) = \sqrt{\frac{6m}{\lambda}} \tanh\left(\sqrt{\frac{m}{2\kappa}}z\right) \label{eq:flatsoliton}
\end{align}
Such a solution interpolates between the values of the order parameter at the two minima of the potential $\mathcal{V}(\psi)$ and has  a $(d-1)$-dimensional, infinite interface centered at $z=0$ with a characteristic thickness $w \sim \sqrt{\kappa/m} $.  What about a solution with a droplet of the $\psi_1$ phase inside a sea of $\psi_2$? Indeed, such solutions are possible if we solve Eq.~\eqref{ELequ} using spherical coordinates.  We would then find a spherically symmetric soliton solution.  

The spherically symmetric case is particularly simple when   the characteristic thickness of $w$   is sufficiently small and  there is a slight asymmetry in the energy of the two minima (a small non-zero $h$, for example). Then,  the spherically symmetric soliton solution has the same form as Eq.~\eqref{eq:flatsoliton}, where we replace $z$ with $r-r_0$, the radial distance away from the droplet surface, taken to be a sphere with radius $r_0$.  This solution, $\psi_I(r-r_0)$, interpolates between the stable phase, $\psi_1$, inside the droplet $(r<r_0)$ and the metastable state $\psi_2$ outside the droplet $(r>r_0)$. This scenario is depicted in Figure \eqref{fig:asympotential}.
Then, substituting this solution into Eq.~\eqref{energyfunc}, we find that the free energy associated with such a droplet is given by 
\begin{align}
 E&=2\pi\kappa \int \dd r\, r \left(\frac{d \psi_I}{d r}\right)^2+2\pi\int \dd r\, r\left(\mathcal{V}(\psi_I)\right) \nonumber\\
&\approx 2\pi r_0\underbrace{\left(\frac{4m\kappa}{\lambda}\sqrt{\frac{m}{2\kappa}}\right)}_{\gamma}-\pi r_0^2\underbrace{(\mathcal{V}(\psi_2)-\mathcal{V}(\psi_1))}_c,
\end{align}
where we have assumed a thin interface width $w \ll r_0$. We have now identified that the free energy has two terms: one proportional to the perimeter $2 \pi r_0$ of the droplet and one to the area $\pi r_0^2$. We may thus identify the associated line tension $\gamma$ and condensation energy $c$.  We have now confirmed that the mean-field solution of this scalar field model is consistent with our phenomenological analysis with the isoperimetric inequality.

What about the effects of thermal fluctuations?  This is a difficult question because the nucleation and growth process is a non-equilibrium phenomenon and using standard equilibrium techniques is problematic. One possibility is to look at this process near the critical droplet size, where we expect the droplet to remain roughly stationary.  In this case, we may suppose that the energy of the droplet in Eq.~\eqref{energyfunc} establishes a Boltzmann distribution with which we may calculate a free energy. Specifically, we may introduce a small field perturbation $\delta \psi(x)$ away from the mean field solution $\psi_I^* \equiv \psi_I(r-r_0^*)$, where the droplet radius is set to the critical value $r_0=r_0^*$.  The free energy, then, would schematically look like
\begin{equation}
\mathcal{F} = - k_BT \ln \int [{\rm d}\delta \psi] \, e^{- \beta \mathcal{H}[\psi_I^*+\delta \psi]}, \label{eq:funcdetmethod}
\end{equation}
where we would integrate over all possible fluctuations $\delta \psi$ away from the critical droplet solution $\psi_I^*$. The first correction amounts to evaluating a functional determinant, as discussed by Callan and  Coleman~\cite{Callan1977}. We will calculate such a correction using a more geometrical approach that treats the undulations of the critical nucleus interface, instead. The connection between these two approaches is discussed in some detail by G\"{u}nther, Nicole, and Wallace \cite{Gunther1980} and Zia \cite{Zia1985}.

\section{Fluctuations near coexistence}

In this section we derive expressions for the line tension of droplets in the plane and on the surface of a sphere including Gaussian fluctuations of the interface away from a midline. Our approach follows the general tack of Zia's analysis \citep{Zia1985}. The derivation ends up being less heuristic in two dimensions, as compared to three dimensions, since the interfaces are curves and the arclength coordinate is easily defined on the entire droplet. Our procedure is:

\begin{enumerate}
\item Construct normal coordinates in the surface near the interface, and in these coordinates compute the metric and gradient operator.
\item Assume that the order parameter, $\psi$, depends only on the coordinate normal to the interface and expand the Euler-Lagrange equations up to $\mathcal{O}\!\left(k^2\right)$ ($k(s)$ being the relevant curvature appearing in the metric).
\item Use these Euler-Lagrange equations to eliminate the potential $\mathcal{V}(\psi(\xi))$ from the total energy $\mathcal{H}$.
\item With this energy compute the partition functions, and thereby the free energy and surface tension.
\end{enumerate}

\subsection{Curves on surfaces}

 Consider a two dimensional surface embedded in three dimensions.  Within this surface lies a curve, the droplet interface, parameterized by its arclength, $\mathbf{R}(s)$. The Darboux frame is constructed from $\hat{\mathbf{t}}$, the unit tangent; $\hat{\pmb{\gamma}}$, the curve normal in the surface; and $\hat{\mathbf{N}}$, the surface normal.  The unit vector $\hat{\pmb{\gamma}}$ is constructed from the other two by the cross-product, $\hat{\pmb{\gamma}}\equiv \hat{\mathbf{N}} \times \hat{\mathbf{t}}$. This frame is determined at every point of the curve by the following set of differential relations:
\begin{align}
\frac{d}{ds} \! \! \begin{pmatrix} \hat{\mathbf{t}}(s) \\ \hat{\pmb{\gamma}}(s) \\ \hat{\mathbf{N}}(s) \end{pmatrix}=\begin{pmatrix}
0 & k_g(s) & k_n(s)  \\ -k_g(s) & 0 & -\tau_g(s) \\ -k_n(s) & \tau_g(s) & 0
\end{pmatrix} \! \! \begin{pmatrix} \hat{\mathbf{t}}(s) \\ \hat{\pmb{\gamma}}(s) \\ \hat{\mathbf{N}}(s) \end{pmatrix},\label{dframe}
\end{align}
where $k_g$ is the geodesic curvature, $k_n$ is the normal curvature, and $\tau_g$ is the geodesic torsion.
We first construct normal coordinates in the vicinity of the curve and within the surface. A point on the surface away from the curve would have a position $\mathbf{r}$ given by
\begin{align}
\mathbf{r}(s,\xi)=\mathbf{R}(s)+\xi\hat{\pmb{\gamma}}(s),
\end{align}
with $\xi$ a distance away from the curve along its normal at an arclength coordinate $s$. Making use of \eqref{dframe} as necessary, we compute the metric:
\begin{align}
g_{ij}&\equiv \frac{d\mathbf{r}}{dq_i}\cdot \frac{d\mathbf{r}}{dq_j}=\begin{pmatrix}
\left(1-\xi k_g\right)^2+\xi^2\tau_g^2 & 0 \\ 0 & 1
\end{pmatrix},
\end{align}
where $q=(s,\xi)$. The derivatives of the order parameter are
\begin{align}
\nabla \psi=&=\frac{\partial \psi}{\partial q_\alpha}g^{\alpha \beta}\frac{\partial \mathbf{r}}{\partial q_\beta}\\
&=\frac{1}{\left(1-\xi k_g\right)^2+(\xi \tau_g)^2}\frac{\partial \psi}{\partial s}\frac{\partial \mathbf{r}}{\partial s}+\frac{\partial \psi}{\partial \xi}\hat{\pmb{\gamma}}\label{grad}.
\end{align}
Now, assuming the order parameter, $\psi(\xi)$, depends on only the normal coordinate $\xi$,  we can write the energy \eqref{energyfunc} as
\begin{align}
\mathcal{H}&=\int \dd s\, \dd \xi \sqrt{\left(1-\xi k_g\right)^2+(\xi \tau_g)^2}\left[\frac{\kappa}{2}(\psi')^2+\mathcal{V}(\psi) \right]. \label{eq:energyintermediate}
\end{align}
The Euler-Lagrange equations \eqref{ELequ} can be used to eleminate the potential from the energy. However the first integral necessary to do so only exists for the flat soliton solution given in Eq.~\eqref{eq:flatsoliton}.  A consistent expansion in the curvature would require knowledge of higher and higher curvature contributions to the Euler-Lagrange equations.  For our purposes,  the lowest order of the Euler-Lagrange equations suffices when considering just one power of the curvature in the energy. However, if one wishes to work at higher orders in the curvature, such as $k_g^2$, more work is required.  We provide an example in the Appendix II.

Since we are going to consider nucleation in the plane and on the sphere, we set $\tau_g(s)=0$ (see the Appendix~I for details). Moreover, we may now set $\psi(\xi,s)$ to the arc-length independent solution  $\psi_0(\xi)$ to the Euler-Lagrange equations. The  energy from Eq.~\eqref{eq:energyintermediate}  then takes the form
\begin{align}
\mathcal{H} \rightarrow \mathcal{H}_{k_g}=\int\!\!\dd s \left( \gamma_0-\gamma_1 k_g(s)\right) \label{Helf}
\end{align}
where $\gamma_0=\kappa\int \psi_0'^2 \dd \xi$ and $\gamma_1=\kappa\int \xi \psi_0'^2 \dd \xi$.  Note that the curvature term proportional to $k_g$ will not contribute for \textit{symmetric} interface profiles $\psi_0(\xi)$ which look the same both in the ordered and disordered phase. This is natural as the sign of the curvature $k_g$ must be defined with respect to either the disordered or ordered phases ({\sl e.g.}, $k_g >0$ for an interface curving into the ordered phase).   Indeed, if there is no difference between the ordered and disordered phases as measured by the distance $\xi$ away from the interface, then this term must be zero by symmetry.

To study the effects of fluctuations of interface we expand the geodesic curvature in small deviations about some reference.  The geodesic curvature can be written as follows
\begin{align}
k_g(s)=\ddot{\mathbf{R}}\cdot(\mathbf{N}\times\dot{\mathbf{R}}).
\end{align}
We are going to expand the curvature in a geodesic polar parametrization,
\begin{align}
r(\phi)=R(1+\epsilon(\phi)),\label{param}
\end{align}
with $\epsilon(\phi)$ describing the small fluctuations around a droplet of radius $R$. We expand $\epsilon(\phi)$ in Fourier modes,
\begin{equation}
\epsilon(\phi)=\sum\limits_{n \neq 0}a_n f_n(\phi)
\end{equation}
where $n$ is an integer and the $f_n(\phi)$ are the set of real orthonormal basis functions (see Appendix~I for details).  Note that $f_0(\phi)$ is a constant so that we may absorb it into the  radius $R$.  So, our summations and products over $n$ in everything that follows are assumed to be over all integers except for zero.

Starting with interfaces in the plane, the normal curvature is zero and the geodesic curvature becomes the only curvature in the problem:
\begin{align}
k_g=k(\phi)=\frac{r^2+2(r')^2-r r''}{\left(r^2+(r')^2\right)^{\frac{3}{2}}}.
\end{align}
Substituting \eqref{param} into the curvature, expanding to quadratic order, and then plugging the result into the curvature energy \eqref{Helf} gives the energy on the plane:
\begin{align}
\mathcal{H}_{\mathbb{R}^2}&=\int  \gamma_0 R\left[ 1+\frac{1}{2}(\epsilon')^2\right] \, \dd \theta -2\pi\gamma_1 \nonumber \\
&    =2 \pi (\gamma_0R-\gamma_1)+\frac{\gamma_0 R}{2} \sum_n n^2 a_n^2.  \label{eq:interfaceE}
\end{align}
Note that the energy in Eq.~\eqref{eq:interfaceE}  is just for the interface of the droplet. In addition, we know that there is a condensation energy that sets the overall size of the droplet. Again, we will be working near critical droplet sizes so that we may treat them as stationary objects with a fluctuating perimeter.

With these developments of the interface description, we now turn to the thermal fluctuations.  An ensemble of critical droplet shapes at some fixed area $A$ and temperature $T$ will have a partition function with a fixed area constraint as follows:
\begin{align}
\mathcal{Z}_{\mathbb{R}^2}=\ell \int  \left[\dd \mathbf{R}\right]\delta\!\left(A-\mathcal{A}[\mathbf{R}]\right)e^{-\beta \mathcal{H}_{\mathbb{R}^2}}
\end{align}
where $\ell$ is a microscopic length scale that will depend on the details of the transition. A similar consideration occurs for fluctuating lipid vesicles, where $\ell$  is of order of a few lipid molecules \cite{faragopincus}. Furthermore, if the transition occurring is one of crystallization, then the relevant length scale will be the lattice spacing. In any case, $\ell$ is at the scale of the (typically microscopic) basic constituents of the system. The measure $[\dd \mathbf{R}]$ represents an integration over all droplet shapes. The area of a particular configuration, $\mathcal{A}[\mathbf{R}]$, in the parametrization \eqref{param} is
\begin{align}
\mathcal{A}[\mathbf{R}]=\pi R^2\left(1+\frac{1}{2\pi}\sum\limits_{n}a_n^2\right)\equiv \pi R^2 (1+\delta_a),
\end{align}
where we have defined $\delta_a$, a convenient representation of the sum in the first equality. The measure, then, may be written as
\begin{align}
\int  \left[\dd \mathbf{R}\right]\equiv \int \mathrm{d}R \int\prod\limits_{n}\dd a_n . \label{eq:measure}
\end{align}
Together with this definition and a change of variables in the delta function, we have
\begin{align}
\mathcal{Z}_{\mathbb{R}^2} & =\ell \int \mathrm{d}R\int\prod\limits_{n \neq 0}^{}\dd a_n \frac{e^{-\beta \mathcal{H}_{\mathbb{R}^2}}}{2 \pi R(1+\delta_a)} \delta\!\left(R-\sqrt{\frac{A/\pi}{1+\delta_a}}\right) \nonumber \\ 
 & =\frac{\ell}{\sqrt{4 \pi A }}\int\prod\limits_{n\neq 0}\dd a_n \frac{e^{-\beta \bar{\mathcal{H}}_{\mathbb{R}^2}}}{\sqrt{1+\delta_a}} ,
\end{align}
where we resolved the delta function by performing the integral over $R$.
 The substitution of $R$ results in a new energy, $\bar{\mathcal{H}}_{\mathbb{R}^2}$, which expanded to quadratic order in $a_n$ is
\begin{align}
\bar{\mathcal{H}}_{\mathbb{R}^2}= \gamma_0\sqrt{4\pi A}-2 \pi\gamma_1+\gamma_0 \sqrt{\frac{A}{4\pi}} \sum_n (n^2-1) a_n^2 \label{energyinftrans}.
\end{align}
The occurrence of $(n^2-1)$ in \eqref{energyinftrans} is not just happenstance: the modes $a_1$ and $a_{-1}$ correspond to infinitesimal translations of the membrane, analogously to the translation modes of three-dimensional vesicles when expanded in spherical harmonics \cite{helfrichves,seifertves}.  Next, we write the prefactor of the Boltzmann weight as follows:
\begin{align}
\frac{1}{\sqrt{1+\delta_a}}&\approx 1-\frac{1}{4\pi}\sum a_n^2 \approx e^{-\frac{1}{4\pi}\sum a_n^2}.
\end{align}

With all of the considerations given thus far the partition function takes the form
\begin{align}
\mathcal{Z}_{\mathbb{R}^2}=\frac{\ell}{\sqrt{4 \pi A }}\int\prod\limits_{n\neq 0}\dd a_n e^{-\frac{1}{4\pi}\sum a_n^2-\beta \bar{\mathcal{H}}_{\mathbb{R}^2}}.
\end{align}
As it stands now, this partition function is problematic for a number of reasons.  First, it does not yield a factor of the area coming from integration over the infinitesimal translations. This occurs because the term in the exponential coming from the delta function ($e^{-\frac{1}{4\pi}\sum a_n^2}$), which we call the Jacobian factor, contains the translation modes.  To deal with this we must also include Faddeev-Popov and Liouville corrections as discussed by Cai {\it et al.} \citep{Cai1994}.  Ordering the calculation by powers of the temperature, the Faddeev-Popov and Liouville corrections come in at the same order as the Jacobian factor.   This is an essential point: to calculate consistently to lowest order, we must neglect the Jacobian term along with these more subtle corrections.
All of these additional corrections are higher order in temperature than the leading-order corrections we are interested in. However, if in the future we wished to study correlation functions, we would be forced to include the other corrections: they are necessary at the outset for a consistent calculation. With these observations in mind, we take our partition function to be
\begin{align}
\mathcal{Z}_{\mathbb{R}^2}=\frac{\ell}{\sqrt{4 \pi A }}\int\prod\limits_{n\neq 0}\dd a_n e^{-\beta \bar{\mathcal{H}}_{\mathbb{R}^2}}.
\end{align}
Let us now go back to the translation modes $a_1$ and $a_{-1}$.  As for analogous analyses of vesicle shape fluctuations \cite{helfrichves,seifertves}, we exclude these modes from our analysis as they contribute to an entropic factor associated with the center of the droplet. Similar zero-energy modes are found in the functional determinant method of calculating the free energy (see discussion around Eq.~\eqref{eq:funcdetmethod}), and such modes must be properly normalized \citep{Coleman1988chp7}.  However, in our simple physical picture of the nucleus as a fluctuating membrane, we can fix the center and ignore these modes entirely, as done for theories of fluctuating vesicles \cite{helfrichves}.

 \begin{figure}
\includegraphics[width=3.5in]{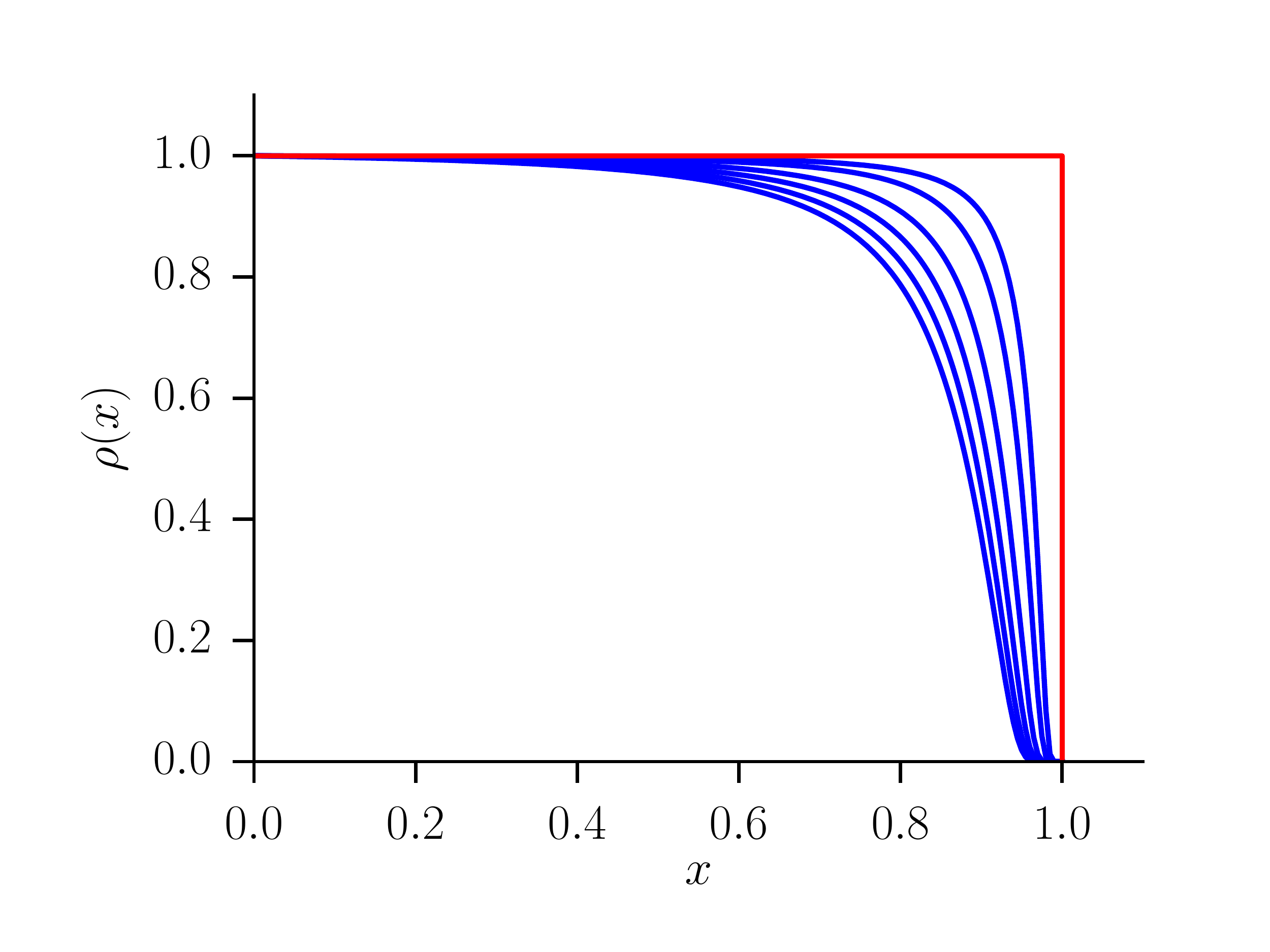}
\caption{A few examples of regularization functions are given. The red square curve corresponds to the hard cutoff defined by $f(x)=1-\Theta[x-1]$. The remaining blue curves are examples of the smooth regularization curves defined as follows: For $0\leq x <1$, $\rho(x)=\exp[ r(-(x-1)^{-2}+1)]$ and for $x\geq 1$, $\rho(x)=0$. The various blue curves have varying $r$ from 0.001 (more sharp) to 0.01 (more smooth).}
\label{fig:regcurves}
\end{figure}

Now we are prepared to compute the free energy, $\beta\mathcal{F}_{\mathbb{R}^2}=-\ln \mathcal{Z}_{\mathbb{R}^2}$,
excluding the translation modes. The rest of the integrations over $a_n$ may be performed as they are all Gaussian integrals. We find
\begin{align}
\beta \mathcal{F}_{\mathbb{R}^2}
& = \ln \left( \frac{P}{\ell}\right)+\beta (\gamma_0 P-2 \pi \gamma_1)\nonumber \\ & \quad {} +\sum\limits_{n>1}\ln \left[\frac{\beta \gamma_0 P  }{4\pi^2}(n^2-1) \right],
\end{align}
where we have recognized that we can replace $A$ with the perimeter $P$ of a disk with the same area $A$, for which $4 \pi A = P^2$. This makes the various terms contributing to the free energy a little more transparent, such as the usual constant line tension term $\beta \gamma_0 P$.  Consider that the first term in the free energy, when given a minus sign and exponentiated, yields the size dependent scaling of the nucleation rate prefactor (recall \eqref{eq:nucrate})  $\Gamma_0\propto 1/P$.  This result matches the results of \citep{Gunther1980, Voloshin1985}, and a requirement for this to be the true scaling is that no term logarithmic in the perimeter can arise from the fluctuation mode sum.  Indeed with a hard cutoff used in the sum, there is no additional logarithm of the perimeter.  What we describe now is a different regularization procedure for which the fluctuation sum does contribute terms logarithmic in the perimeter.

To complete these sums we a introduce a cutoff function $\rho(n/N)$ where $\rho(0)=1$ and compactly supported.  Consider $\rho(x)$ as a smoothed and compactly supported version of the function $f(x)=1-\Theta[x-1]$, where $\Theta[x]$ is the Heaviside step function. Note that $f(x)$ is the usual hard cutoff at $n=N$. Such an example is depicted in Figure \eqref{fig:regcurves}. The regularity of the cutoff function is related to the asymptotic estimates for the sums: more regularity leads to better estimates in powers of $1/N$. For an accessible review and introduction to these methods, see Tao's online notes \citep{Tao2010}. Within this regularization scheme, the free energy becomes (for $N \gg 1$):
\begin{align}
\beta\mathcal{F}_{\mathbb{R}^2}&\approx \beta \gamma P \left(1-\frac{2 \pi \delta_T}{P}\right) + \ln\left( \frac{\pi P}{\ell}\right)\nonumber \\&\quad {} +\left(-\frac{3}{2}+ N \!\braket{\rho}\right) \ln\left(\frac{\beta \gamma P }{4 \pi^2} \right) \nonumber \\ & {} \quad +N \int_0^1 \ln x^2 \rho(x) \, \mathrm{d} x+ N \ln N^2 \langle \rho \rangle  \label{eq:flatFE}\\
& \approx\beta \gamma P \left(1-\frac{2 \pi \delta_T}{P}\right) + \ln\left( \frac{\pi P}{\ell}\right)-\frac{3}{2}\ln\left(\frac{\beta \gamma P }{4 \pi^2} \right) \nonumber \\&\quad {} +N\left[\ln\left(\frac{\beta \gamma P }{4 \pi^2} \right) -2+  \ln N^2  \right]. \label{eq:flatFE2}
\end{align}
where $\langle \rho \rangle \equiv \int_0^1\rho(x)\,\mathrm{d}x \approx 1$, and we redefined $\gamma_0=\gamma$, $\gamma_1=\gamma \delta_T$.

Equations~\eqref{eq:flatFE}, \eqref{eq:flatFE2} are our main result of the thermal fluctuation analysis. The first term proportional to $\beta$ is the usual mean-field result, which includes a Tolman length $\delta_T$. Note that the Tolman length $\delta_T$ is proportional to $\gamma_1$, which we argued vanishes for symmetric phases, which has also been argued on general grounds for the three-dimensional problem \cite{Fisher1984}.  Assuming the highest accessible undulation mode is given by $N=P/\ell$, where, again, $\ell$ is the length scale of the microscopic constituents of the droplet, we see that the nucleation rate prefactor takes the form $\Gamma_0\propto P^{1/2}$.
It is for this reason that we discuss the regularization dependence on the nucleation rate. Note that G\"{u}nther {\sl et al}. \citep{Gunther1980} identify the three-dimensional case as a special one because the equivalent of our fluctuating sum contributes non-trivially to the nucleation rate (and does not for two and four dimensions).

 Note that the contribution to $\Gamma_0$ comes from the finite piece of the sum over modes $n$, which evidently depends on the choice of regularization procedure. It's possible that our model would require modifications to the integral measure in Eq.~\eqref{eq:measure} to remove this dependence, as the issue of having an infinite number of modes $a_n$ already appears in this measure. These subtleties may be the source of discrepancy between our calculation and previous work such as Garriga's calculation \citep{Garriga1994}, done using zeta-function regularization, and Voloshin's analysis \citep{Voloshin1985}, which connects the integration over modes $a_n$ to the harmonic oscillator partition function via the path integral. These and other studies   ended up yielding the $\Gamma_0 \propto 1/ P$ result we find with a hard cutoff. An additional subtlety is that these previous works do not fix the droplet area $A$   and deal with the resulting unstable mode (the $a_0$ mode which gets a negative eigenvalue) using analytic continuation.   In our analysis, by fixing the droplet area, we do not have to worry about the unstable $a_0$ mode.  Reconciling these approaches would be an interesting topic for further study.

Unlike the $\Gamma_0$ contributions, both the smooth and hard cutoffs yield the same divergent terms (terms proportional to $N$ in \eqref{eq:flatFE2}), which we will study in more detail below. Indeed, aside from the consideration of the nucleation rate, we may treat the various  fluctuation corrections in Eq.~\eqref{eq:flatFE2} as renormalizations of the line tension $\gamma$.  These terms will introduce  weak, logarithmic dependence on the perimeter $P$.  For reasonable parameter values, these corrections serve to \textit{increase} the effective line tension compared to the bare value $\gamma$.  This is analogous to the correction to the surface tension in two-dimensional membranes \cite{davidleibler}.   

We  note that interest in these logarithmic corrections is not new.  Schmitz {\it et al.} have studied the impact of interface fluctuations on logarithmic contributions to the surface tension \citep{Schmitz2014}.  As previously noted, the interest in logarithmic corrections as it pertains to scaling in nucleation rates spans a wide range of fields: from cosmology to chemical physics.\citep{Coleman1977,Callan1977,Langer1967,Gunther1980,Binder2016}  These same corrections appear throughout the membrane literature as thermal corrections to membrane moduli. \citep{Morse1995,Peliti1985,Cai1994,davidleibler} Most of this work focuses on three dimensions, and in terms of the prefactor to the nucleation rate, a general consensus is that $\Gamma_0 \propto A^{7/6}$, where $A$ is the surface area of the nucleated droplet \citep{Garriga1994,Gunther1980,Prestipino2014}.
 
\begin{figure}
\includegraphics[width=2.1in]{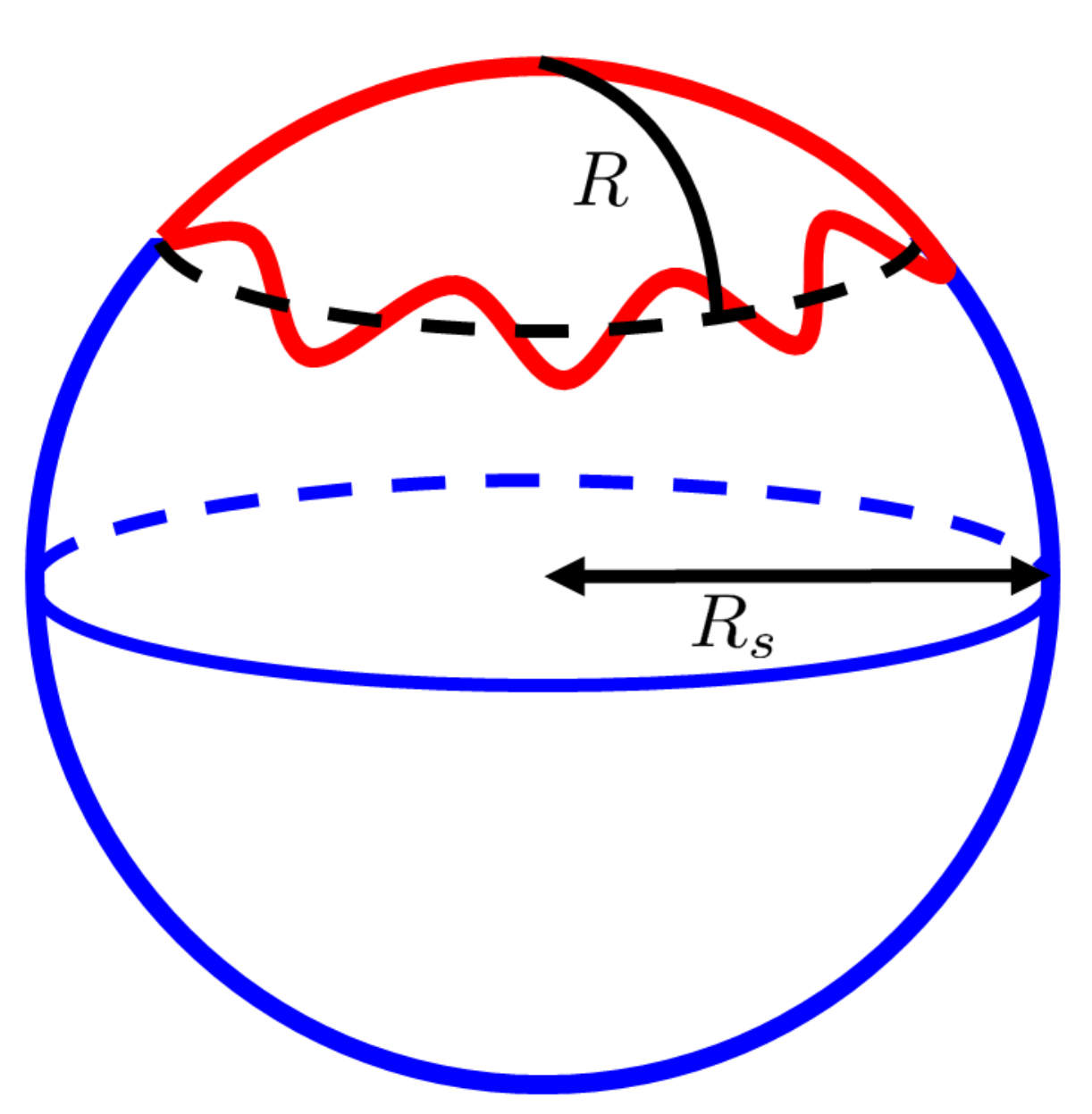}
\caption{The red curve is a schematic of the fluctuation interface of a droplet on the sphere.}
\label{fig:spherewobble}
\end{figure}

In the same fashion one may calculate the free energy for nucleation on the surface of a sphere of radius $R_s$. The curve on the sphere is parametrized in geodesic polar coordinates $r(\phi)=R(1+\epsilon(\phi))$, as shown in Fig.~\ref{fig:spherewobble}. The quantity $R$ is the geodesic distance from the center of the droplet to the midline of the droplet edge, analogous to the the radius parameter $R$ used in the flat case. The process for analyzing the fluctuations is the same as in the plane, including the integration over $R$. The resulting free energy reads, for large $N$,\begin{align}
\beta \mathcal{F}_{S^2}&\approx\beta \gamma P\left[1- \frac{2 \pi\delta_T}{P}(1-2\bar{a})\right]+\ln \left(\frac{\pi P}{\ell}\right)\nonumber\\
&\quad{}+\left(-\frac{3}{2}+N\langle \rho \rangle\right)\ln\left[\frac{\beta\gamma    R_s ^2(\cos^{-1}(1-2\bar{a}))^2}{P}\right] \nonumber \\
& \quad {}+N \int \ln x^2 \rho(x) \, \mathrm{d} x+ N \ln N^2 \langle \rho \rangle \\
&\approx\beta \gamma P\left[1- \frac{2 \pi\delta_T}{P}(1-2\bar{a})\right]+\ln \left(\frac{\pi P}{\ell}\right)\nonumber\\
&\quad{}+\left(-\frac{3}{2}+N \right)\ln\left[\frac{\beta\gamma    R_s ^2(\cos^{-1}(1-2\bar{a}))^2}{P}\right] \nonumber \\
& \quad {}-2N+ N \ln N^2, 
\end{align}
where $\bar{a}=A/4\pi R_s^2$ is a rescaled area of the nucleated region, and we still have  $\gamma_0=\gamma$ and $ \delta_T=\gamma_1/\gamma$. We wrote our expression in terms of the perimeter $P$ of a geodesic disk with area $A$, which satisfies $P=4 \pi R_s \sqrt{\bar{a}(1-\bar{a})}$.  Note an interesting property of the nuclei on the sphere: When $\bar{a}=1/2$, the Tolman length term vanishes, regardless of the value of the Tolman length! This makes intuitive sense because when $A = 2 \pi R_s^2$, the nucleus fills a hemisphere.  Therefore, its boundary is on the equator of the sphere and is a perfectly straight interface! Indeed, unlike the plane, it is possible to have a finite-sized region on the sphere with a perfectly straight boundary.

\begin{figure}
\includegraphics[width=\linewidth]{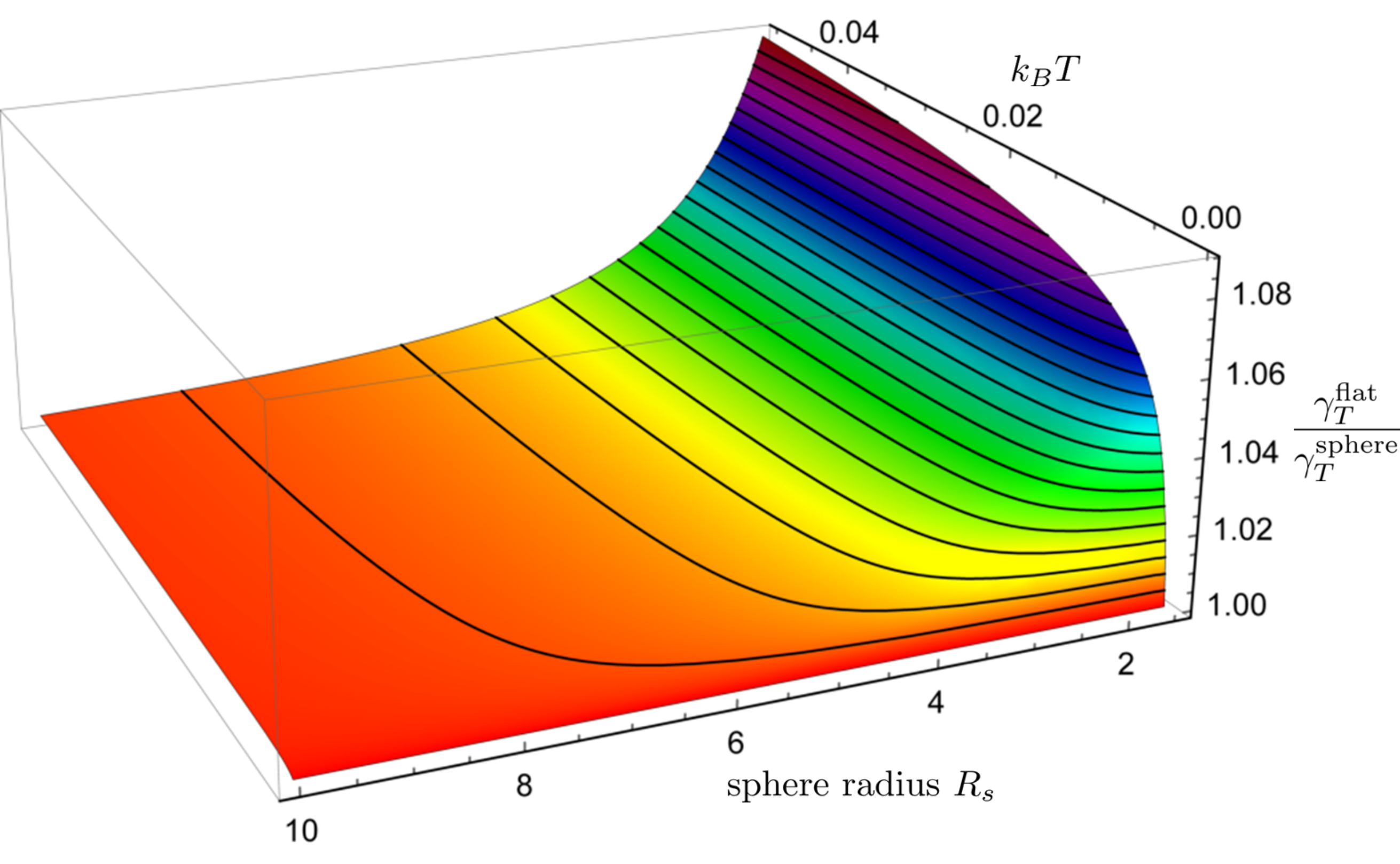}
\caption{We compare the effective surface tension, renormalized by temperature fluctuations, between the flat and spherical case. The flat surface tension is always larger, as shown by the plotted ratio $\gamma_T^{\mathrm{flat}}/\gamma_T^{\mathrm{sphere}} > 1$. We compared the tensions at the critical droplet sizes for the two respective cases.  We used the bare tension value $\gamma=2$, a condensation energy $c=1$, and $\ell=0.01$.  As expected, the two tensions coincide as $R_s$ becomes large.  Also, as $T \rightarrow 0$, both tensions will just reduce to their bare value of $\gamma$.   }
\label{fig:effectivetension}
\end{figure}

Apart from the difference in the Tolman term, the sphere energy has a different fluctuation correction proportional to $N$. One can verify that for small areas $A \ll 4 \pi R_s^2$, that the two corrections coincide in the flat and sphere case.  It's interesting to compare the renormalized line tensions in the two cases. Setting $N = P/\ell$, we find that the renormalized tension $\gamma_T\equiv \gamma+\delta \gamma$ has 
\begin{align}
\delta \gamma= \begin{cases}
\dfrac{k_BT}{\ell}\ln\left[\dfrac{\beta\gamma  P^3}{(2\pi e \ell )^2}\right] &\mbox{flat} \\[15pt]
\dfrac{k_BT}{\ell}\ln\left[\dfrac{\beta    \gamma P R_s ^2(\cos^{-1}(1-2\bar{a}))^2 }{(e \ell)^2}\right] &\mbox{sphere}
\end{cases}.
\end{align}
The tensions are  similar. Note that we have to evaluate these tensions at the respective critical droplet sizes for the two cases. This is because, as discussed above, the free energy approach only makes sense near the critical droplet size where the system may be approximated as stationary, with a Boltzmann distribution over fluctuating interface modes. 

  We can look at the difference between the spherical and flat case by looking at the ratio $\gamma_T^{\mathrm{flat}}/\gamma_T^{\mathrm{sphere}}$, which is plotted in Fig.~\ref{fig:effectivetension}. The difference between the two for reasonable parameter values is of order 10\% for the largest values.  We see that the surface tension is always larger in the flat case. The behavior of the ratio is understandable. First, as $T \rightarrow 0$, both of the tensions go to their bare values $\gamma$, so that the ratio goes to one.  Similarly, as $R$ increases, the spherical case must approach the flat one and we also get a ratio approach unity. The crossover lengthscale will be related to the critical droplet radius, which for the choice of parameters given in Fig.~\ref{fig:effectivetension}  is given by $R^*= \gamma/c=2$. So, as the sphere radius $R$ becomes much larger than this critical droplet size, the flat and sphere case begin to coincide. This is evident from Fig.~\ref{fig:effectivetension}.

\section{Conclusions}
We established a mean-field result for the critical nucleus from an isoperimetric inequality. This inequality provides the intuition that nucleation on surfaces of positive Gaussian curvature is easier and harder on negative Gaussian curvature surfaces compared to the flat  space case. We then considered a more proper model of nucleation which includes a  spatially varying order parameter that describes the transition region between the two phases. We showed that the scaling of the nucleation rate prefactor may be sensitive to the precise way in which the nucleating droplet is described and may not be as universal as currently believed. Furthermore, we focused on comparing nucleation on a sphere versus the flat plane, taking into account thermal fluctuations. We found that the fluctuations serve to increase the effective line tension of the critical nucleus in both the sphere and flat cases. We also found that the thermal fluctuation renormalizations are less severe in the spherical case.

The number of examples of nucleation phenomena occuring in 2D is proliferating at a considerable rate.  Our result on the nucleation rate prefactor is a testable prediction, and we would like to find an experiment to make that test.  In the near future there is the possibility to numerically study the Ising model, which is known to obey the classical nucleation theory quite well \cite{IsingCNT}. An extension of a functioning simulation of the Ising model critical droplets in 2D would be to do the simulation on a sphere.

\acknowledgments
 This work was supported in part through NSF DMR12-62047.  This work was partially supported by a Simons Investigator award from the
Simons Foundation to Randall D.~Kamien. M.O.L. gratefully acknowledges partial funding from the Neutron Sciences Directorate (Oak Ridge National Laboratory), sponsored by the U.S. Department of Energy, Office of Basic Energy Sciences.

\section{Appendix I}

When expanding the droplet interface around its midline by introducing a small deviation $\epsilon(\theta)$, it is convenient to use the following set of functions as the basis for expansions in Fourier series:
\begin{flalign}
\underline{n>0}:      && f_n(\theta)&= \frac{\sin(n\theta)}{\sqrt{\pi}}&&\nonumber\\
\underline{n<0}: && f_n(\theta)&= \frac{\cos(n\theta)}{\sqrt{\pi}},&&\nonumber
\end{flalign}
where $\theta \in [0,2\pi)$ parameterizes the droplet edge. With these definitions the basis is orthonormal. Furthermore, such definitions will make the functional integral simpler. A general expansion for $\epsilon(\theta)$  now takes the form
\begin{align}
\epsilon=\sum\limits_{|n|>0}a_n f_n
\end{align}
so,
\begin{align}
&\int \dd\theta \,\epsilon^2=\sum\limits_{|n|>0}a_n^2\\
&\int \dd\theta \,(\epsilon')^2=\sum\limits_{|n|>0}n^2a_n^2.
\end{align}

We now consider the shape of the midline of the droplet on the surface of a sphere. The geodesic torsion is defined as follows
\begin{align}
\tau_g(s)\equiv \frac{d{\hat{\mathbf{N}}}}{d s}\cdot\hat{\pmb{\gamma}}
\end{align}
If we are on a sphere of radius $R_s$,
\begin{align}
 \frac{d{\hat{\mathbf{N}}}}{d s}\cdot\hat{\pmb{\gamma}}=\frac{1}{R_s}\frac{d{\mathbf{R}}}{d s}\cdot\hat{\pmb{\gamma}}=\frac{1}{R_s} \,\hat{\mathbf{t}}\cdot\hat{\pmb{\gamma}}=0.
\end{align}
The geodesic torsion is also zero in the plane since $\hat{\mathbf{N}}$ is constant.
The geodesic curvature takes the form
\begin{align}
k_g(s)=\ddot{\mathbf{R}}\cdot(\mathbf{N}\times\dot{\mathbf{R}}).
\end{align}
To make progress we need to convert the m{\'e}lange of vectors as functions of the arclength into a function of $\phi$ (the azimuthal angle on the sphere). Note that a dot indicates a derivative with respect to arclength while a prime  indicates a derivative with respect to the angle $\phi$.
\begin{align}
\frac{d\mathbf{R}}{ds}&=\left(\frac{ds}{d\phi}\right)^{-1}\frac{d\mathbf{R}}{d\phi}\\
\frac{d^2\mathbf{R}}{ds^2}&=\left(\frac{ds}{d\phi}\right)^{-1}\frac{d}{d\phi}\left(\left(\frac{ds}{d\phi}\right)^{-1}\frac{d\mathbf{R}}{d\phi}\right)\\
&=-\left(\frac{ds}{d\phi}\right)^{-3}\frac{d^2s}{d\phi^2}\frac{d\mathbf{R}}{d\phi}+\left(\frac{ds}{d\phi}\right)^{-2}\frac{d^2\mathbf{R}}{d\phi^2}\\
&=-\left(\frac{ds}{d\phi}\right)^{-2}\frac{d^2s}{d\phi^2}\mathbf{t}+\left(\frac{ds}{d\phi}\right)^{-2}\frac{d^2\mathbf{R}}{d\phi^2}
\end{align}
We can parametrize the spherical curve by
\begin{align}
\mathbf{R}=R_s\mathbf{N}=R_s\begin{pmatrix}
\cos\phi \sin\theta \\
\sin\phi \sin\theta \\
\cos\theta
\end{pmatrix}.
\end{align}
The arclength obeys the differential equation
\begin{align}
\frac{ds}{d\phi}=R_s\sqrt{\sin^2\theta+\left(\frac{d\theta}{d\phi}\right)^2}.
\end{align}
Now we take the necessary derivatives of $\mathbf{R}$ and then substitute into the expression for the geodesic curvature.
\begin{align}
k_g(\phi)=\ddot{\mathbf{R}}\cdot(\mathbf{N}\times\dot{\mathbf{R}})=\left(\frac{ds}{d\phi}\right)^{-3}\mathbf{R}''\cdot(\mathbf{N}\times\mathbf{R}')
\end{align}
A short calculation gives
\begin{align}
\mathbf{R}''\cdot(\mathbf{N}\times\mathbf{R}')=&R_s^2\left(2\cos\theta \left(\frac{d\theta}{d\phi}\right)^2\right.\\
+&\left.\sin^2\theta \cos\theta - \sin\theta \frac{d^2\theta}{d\phi^2}\right).
\end{align}
The geodesic curvature then reads
\begin{align}
k_g=\frac{2\cos\theta \left(\frac{d\theta}{d\phi}\right)^2+\sin^2\theta \cos\theta - \sin\theta \frac{d^2\theta}{d\phi^2}}{R_s\left(\sin^2\theta+\left(\frac{d\theta}{d\phi}\right)^2\right)^{3/2}}.
\end{align}
As a check, we can take two limits to verify the expression for the geodesic curvature.  In the first, we assume there is no variation in the interface, looking to find the constant geodesic curvature expression for circles on the sphere.
\begin{align}
\text{circle: } k_g=\frac{1}{R_s}\frac{\sin^2\theta \cos\theta}{\sin^3\theta}=\frac{1}{R_s\tan\theta}
\end{align}
If we let $\theta=r/R_s$, where $r$ is a geodesic polar radius, and let $R_s\rightarrow \infty$
\begin{align}
\lim\limits_{R_s\rightarrow \infty}k_g=\lim\limits_{R_s\rightarrow \infty}\frac{1}{r+\mathcal{O}(\frac{1}{R_s^2})}=\frac{1}{r}.
\end{align}
The curvature of a circle of radius $r$ in the plane!
Now let us put the derivatives back but with the substitution $\theta=r/R_s$.
\begin{align}
k_g
&=\frac{2\cos \left(\frac{r}{R_s}\right)\left[ \left(\frac{dr}{d\phi}\right)^2+\frac{R_s^2 }{2}\sin^2\left(\frac{r}{R_s}\right) \right]- R_s \sin\left(\frac{r}{R_s}\right) \frac{d^2r}{d\phi^2}}{\left[ R_s^2\sin^2\left(\frac{r}{R_s}\right)+\left(\frac{dr}{d\phi}\right)^2\right]^{3/2}}
\end{align}
When we take $R_s\rightarrow \infty$,
\begin{align}
\lim\limits_{R\rightarrow \infty} k_g&=\lim\limits_{R\rightarrow \infty}\frac{2 \left(\frac{dr}{d\phi}\right)^2+r^2- r \frac{d^2r}{d\phi^2}+\mathcal{O}(\frac{1}{R_s^2})}{\left(r^2+\left(\frac{dr}{d\phi}\right)^2+\mathcal{O}(\frac{1}{R_s^2})\right)^{3/2}}\nonumber\\
&=\frac{2(r')^2+r^2-r r''}{\left(r^2+(r')^2\right)^{3/2}}
\end{align}
This is exactly the expression for the curvature in polar coordinates in the plane.

\section{Appendix II \label{appx3}}
Here we establish that the infinite interface form for the order parameter is the correct substitution to order $k^2$. We work in the spirit of the calculation of Ref.~\citep{Fisher1984}.

Begin with the following functional:
\begin{align}
\mathcal{H}=\int \dd^2x \left(-\frac{\kappa}{2}\psi \nabla^2\psi+\frac{c}{2}\left(\nabla^2\psi\right)^2 +\mathcal{V}(\psi)\right)
\end{align}
The extremizer of this functional is the solution to the following Euler-Lagrange equation:
\begin{align}
-\kappa\nabla^2 \psi+c\nabla^4 \psi =-\frac{\partial \mathcal{V}}{\partial \psi} \label{eq:appxEL}
\end{align}
In the normal coordinates used throughout these notes, the Laplacian takes the form
\begin{align}
\nabla^2\psi=\frac{\partial^2 \psi}{\partial \xi^2}-\frac{k_g(s)}{\left(1-\xi k_g(s)\right)}\frac{\partial \psi}{\partial \xi}
\end{align}
where the order parameter, $\psi(\xi)$, takes a particular form:
\begin{align}
\psi(s,\xi)=\psi_0(\xi)+k(s)\psi_1(\xi)+k(s)^2\psi_2(\xi)
\end{align}
If we apply the Laplacian again to get the 4th derivative terms, then we will generate derivatives of the geodesic curvature with respect to the arclength.  To simplify the problem we assume that $k_g(s)$ is slowly varying and, therefore, the derivatives can be set to zero.

Given below are expansions in powers of the curvature for various quantities appearing in the energy functional.

\begin{align}
\nabla^2 \psi\approx\psi_0''+k\left(\psi_1''-\psi_0'\right)+k^2\left(\psi_2''-\psi_1'-\xi\psi_0'\right)
\end{align}
\begin{align}
\nabla^4\psi\approx&\psi_0^{(4)}+k\left(\psi_1^{(4)}-2\psi_0^{(3)}\right)\nonumber\\
&+k^2\left(\psi_2^{(4)}-2\psi_1^{(3)}-2\xi\psi_0^{(3)}-\psi_0''\right)
\end{align}
\begin{align}
\mathcal{V}(\psi)=&\mathcal{V}(\psi_0)+k\psi_1\left.\frac{\partial\mathcal{V}}{\partial \psi}\right|_{\psi_0}\nonumber\\
&+k^2\left(\psi_2\left.\frac{\partial\mathcal{V}}{\partial \psi}\right|_{\psi_0}+\frac{1}{2}\psi_1^2\left.\frac{\partial^2 \mathcal{V}}{\partial\psi^2}\right|_{\psi_0}\right) 
\end{align}
With these results in hand, we expand the Euler-Lagrange equation \eqref{eq:appxEL} in powers of the curvature $k$:
\begin{align}
\begin{cases} 
      -\kappa \psi_0''+c\psi_0^{(4)}+\left.\frac{\partial \mathcal{V}}{\partial \psi}\right|_{\psi_0}=0 & k^0 \\
      -\kappa\left(\psi_1''-\psi_0'\right)+c\left(\psi_1^{(4)}-2\psi_0^{(3)}\right)+\psi_1\left.\frac{\partial^2\mathcal{V}}{\partial \psi^2}\right|_{\psi_0}=0 & k^1 
   \end{cases}
\end{align}
There is a      ``first integral" for the $k^0$-order Euler-Lagrange equation:
\begin{align}
\psi_0'\left(-\kappa \psi_0''+c\psi_0^{(4)}+\left.\frac{\partial \mathcal{V}}{\partial \psi}\right|_{\psi_0}\right)=0,\\
\Rightarrow \left(-\frac{\kappa}{2}(\psi_0')^2+c\psi_0'\psi_0'''-\frac{c}{2}(\psi_0'')^2+\mathcal{V}\right)'=0
\end{align}
which is:
\begin{align}
\left(-\frac{\kappa}{2}(\psi_0')^2+c\psi_0'\psi_0'''-\frac{c}{2}(\psi_0'')^2+\mathcal{V}\right)=A
\end{align}
The constant $A$ is fixed by enforcing the boundary conditions. As $\xi\rightarrow -\infty$, $\psi'(\xi)\rightarrow 0$ (the reason for considering the metastable state as being at $\xi\rightarrow -\infty$ is from the construction of the normal to the curve in the surface, which results in $\xi>0$ inside the droplet) and we get
\begin{align}
\mathcal{V}\left(\psi(-\infty)\right)=A
\end{align}
The order parameter at infinity takes on the value corresponding to the metastable state.  We can, without loss of generality, set the value of the potential at the metastable state to be zero.
Now we are free to write the first integral as
\begin{align}
-\frac{\kappa}{2}(\psi_0')^2+c(\psi_0'\psi_0'')'-c(\psi_0'')^2-\frac{c}{2}(\psi_0'')^2+\mathcal{V}(\psi_0)=0. \label{eq:appxFI}
\end{align}

The energy functional as series in the curvature has a leading-order term given by 
\begin{align}
\mathcal{H}_{k^0}=\int\dd s \,  \dd \xi \left(\frac{1}{2}\kappa(\psi_0')^2+\frac{1}{2}c(\psi_0'')^2+\mathcal{V}(\psi_0)\right)
\end{align}
From the first integral of motion, we have
\begin{align}
\mathcal{V}(\psi_0)=\frac{\kappa}{2}(\psi_0')^2-c(\psi_0'\psi_0'')'+\frac{3c}{2}(\psi_0'')^2.
\end{align}
Now we can rewrite this leading order term as
\begin{align}
\mathcal{H}_{2,k^0}=\int\dd s \,  \dd \xi \left(\kappa(\psi_0')^2+2c(\psi_0'')^2\right)
\end{align}
The next order in the curvature $k$ yields the contribution
\begin{align}
\mathcal{H}_{k^1}=\int\dd s \,  \dd \xi&\left[\psi_1\left.\frac{\partial\mathcal{V}}{\partial \psi}\right|_{\psi_0}+\kappa \psi_0'\psi_1'-c\psi_0'\psi_0''+c\psi_0''\psi_1''\right.\nonumber\\
&\left.-\xi\left(\mathcal{V}(\psi_0)+\frac{1}{2}\kappa(\psi_0')^2+\frac{1}{2}c(\psi_0'')^2\right)\right]k(s)
\end{align}
By substituting in for the zeroth order solution $\psi_0$ to the Euler-Lagrange equation, and using the first integral in \eqref{eq:appxFI}, we rewrite this contribution as
\begin{align}
\mathcal{H}_{k^1}=\int\dd s  \,  \dd \xi  \,  \xi\left(\kappa(\psi_0')^2+2c(\psi_0'')^2\right)k(s).
\end{align}
Finally, at order $k^2$, we find the contribution
\begin{align}
\mathcal{H}_{k^2}=\frac{1}{2}\int\!\dd s\, \dd \xi \, f[\xi,{\psi_i(\xi)}]\,k^2(s),
\end{align}
where
\begin{align}
f[\xi,\{\psi_i\}]=&(-2\xi\psi_1+2\psi_2)\left.\frac{\partial\mathcal{V}}{\partial \psi}\right|_{\psi_0}+\psi_1^2\left.\frac{\partial^2\mathcal{V}}{\partial \psi^2}\right|_{\psi_0}\nonumber\\
&+c(\psi_0')^2-2\xi\kappa\psi_0'\psi_1'+\kappa(\psi_1')^2\nonumber\\
&-2c\psi_1'\psi_0''-2c\psi_0'\psi_1''-2c\xi\psi_0''\psi_1'' \nonumber\\
&+c(\psi_1'')^2+2c\psi_0''\psi_2''+2\kappa\psi_0'\psi_2'.
\end{align}
This contribution may be simplified by using the Euler-Lagrange equations and integration by parts. The simplified expression reads
\begin{align}
\mathcal{H}_{k^2}=\frac{1}{2}\int\dd s \, \dd \xi\left(\kappa\psi_1\psi_0'+2c\psi_1'\psi_0''+c(\psi_0')^2\right)k^2(s).
\end{align}
The correction $\psi_1$ to the Euler-Lagrange equation starts to come in at order $k^2$ in the energy.
\bibliographystyle{abbrv}
\bibliography{thebibs}{}

\end{document}